\documentclass[aps,amsfonts,prl,showpacs,nobibnotes,nofootinbib,%
tightenlines]{revtex4}
\usepackage{amsmath,amssymb}
\input BoxedEPS
 \SetRokickiEPSFSpecial  
\HideDisplacementBoxes

\newcommand{\beq}{\begin{equation}}
\newcommand{\eeq}{\end{equation}}
\newcommand{\ba}{\begin{array}}
\newcommand{\ea}{\end{array}}

\begin{document}

\title{Comment on hep-ph/0404212 by D.~Diakonov, V.~Petrov, and
M.~Polyakov}
\author{R.~L.~Jaffe}

\affiliation{Center for Theoretical Physics, Laboratory for Nuclear
Science and Department of Physics, Massachusetts Institute of Technology,
Cambridge, Massachusetts 02139\\
MIT-CTP-3498}

\begin{abstract}\noindent

In hep-ph/0404212 Diakonov, Petrov, and Polyakov respond to my recent
comments on their 1997 paper, ``Exotic anti-decuplet of baryons:
prediction from chiral solitons''.  Their responses do not address
the basic issues or alter the conclusions in my paper.
\end{abstract}
\pacs{12.38.-t, 12.39.Dc, 12.39.-x, 14.20-c}
\vspace*{-\bigskipamount}

\preprint{MIT-CTP-3498}
\maketitle

In January of this year I posted a comment\cite{rlj} on the 1997
paper\cite{dpp} in which Diakonov, Petrov, and Polyakov (DPP) study
the exotic baryon antidecuplet in the chiral soliton model (CSM).
Ref.~\cite{dpp} is an important paper.  In it DPP argue that the
existence of a light, narrow, $S=+1$ exotic baryon is a robust
prediction of the CSM. The recent reports of such a state
($\Theta^{+}(1530)$) seem to confirm this
prediction\cite{Nakano:2003qx}.  Anyone interested in this subject
will want to read Ref.~\cite{dpp} carefully.

The principal point of my comment was that the calculation of the
width of the $\Theta$ presented in Ref.~\cite{dpp} contains an
arithmetic mistake.  When the error is corrected, the result of the
calculation is $\Gamma(\Theta^{+})\approx$ 30 MeV, not 15 MeV as
reported in Ref.~\cite{dpp}.  In passing I also pointed out a) that an
earlier paper\cite{DP} frequently cited by DPP as the first discussion
of an exotic antidecuplet in the CSM actually does not mention that
set of states at all.  Instead the antidecuplet appears to have been
first mentioned in this context by Manohar\cite{Manohar}, and shortly
afterwards by Chemtob\cite{Chemtob}, and Biedenharn and
Dothan\cite{BandD}; and b) that a mass of 1530 MeV for the $\Theta^+$
was computed in the Skyrme model --- a particular version of the CSM ---
by Praszalowicz in 1987\cite{mp}.

In April DPP posted a note in which they respond to my
statements\cite{dpp2}.  Nothing in Ref.~\cite{dpp2} in any way
contradicts points a) and b).\footnote{In Version 2 of their note they
list two other early calculations of the $\Theta$ mass which give
values close to the one obtained by Praszalowicz.} On the more
important issue of the width calculation, they put forward two new and
different arguments, not to be found anywhere in their 1997 paper, to
justify a value of 15 MeV for the width of the $\Theta^{+}$.  One
starts from the experimentally measured value of $g_{\pi N\!N }$.  The
other replaces a crucial factor of $M_{N}/M_{\Delta}$ by
$M_{\Delta}/M_{N}$.  In Ref.\cite{dpp2} DPP treat the
spin-3/2 decay widths as if they had been only secondary in their 1997
paper.  However, careful reading of the 1997 paper clearly shows that
the spin-3/2 decay widths were in fact used to normalize all matrix
elements.  In any case, while the new arguments should be judged on
their own merits, they do not bear on the question raised in my
comment, which concerned what was written and calculated in the 1997
paper.

My persistence in this matter has been fuelled principally by an
e-mail exchange between M.~Polyakov and H.~Weigel which took place in
1998.  In it Weigel asks about the same apparent inconsistency in
Ref.~\cite{dpp} that I discuss in Ref.~\cite{rlj}\footnote{In fact I
first learned of a problem in Ref.~\cite{dpp} in a footnote in
Weigel's 1998 paper on exotics in the CSM\cite{hw}}.  In his reply to
Weigel Polyakov clearly and directly admits that he made an arithmetic
error in the calculation in question.  Since Polyakov apparently was
responsible for the calculation and explicitly admits the only point
at issue, it seems unnecessary to continue this discussion further.

In their recent note Diakonov et al misrepresent the content of that
email exchange.  Because of its importance I have asked Weigel, the
recipient of the e-mail, for permission to make a copy of the original
email publicly available.  The reader may find it at $\ \langle$
http://pierre.mit.edu/$\sim$jaffe/pwemail.html$\ \rangle$.  The first
paragraph of Polyakov's reply is the relevant material.  His later
remarks, although interesting in retrospect, do not bear on the
question of an error in Ref.~\cite{dpp}.  The basic documents in this
controversy are now all public (Refs.~\cite{dpp,rlj,dpp2} and the
Polyakov-Weigel email) and the interested reader can evaluate the
situation for him/her self.
 
I thank H.~Weigel for permission to make his email publicly
available, and F.~Close and H.~Weigel for helpful discussions and
comments on this manuscript.  This work is supported in part by the
U.S.~Department of Energy (D.O.E.) under cooperative research
agreement~\#DF-FC02-94ER40818.



\end{document}